\nonstopmode

\documentclass[prb,showpacs,twocolumn]{revtex4}
\usepackage{graphicx}
\usepackage{subfigure}
\usepackage{color}

\def\jlt#1{{J. Lightwave Technol.} \textbf{#1}}
\def\jqe#1{{IEEE J. Quantum Electron.} \textbf{#1}}

\def\ptl#1{{IEEE Photon.\ Technol.\ Lett.} \textbf{#1}}
\def\opex{ Opt.\ Express }
\def\ao{ Appl.\  Opt.\ }

\def\apl{ Appl.\ Phys.\ Lett.\ }

\def\jqe{ IEEE J.\ Quantum Electron.\ }

\def\ol{ Opt.\ Lett.\ }

\def\pra{ Phys.\ Rev.\ A }

\begin{document}
\title{Stimulated Raman scattering cascade spanning the wavelength range of 523 to 1750~nm using a graded-index multimode optical fiber}
\author{Hamed Pourbeyram}
\affiliation{Department of Electrical Engineering and Computer Science, University of Wisconsin-Milwaukee, Milwaukee, WI 53211, USA}
\author{Govind P. Agrawal}
\affiliation{The Institute of Optics, University of Rochester, Rochester, NY 14627, USA}
\author{Arash Mafi}
\affiliation{Department of Electrical Engineering and Computer Science, University of Wisconsin-Milwaukee, Milwaukee, WI 53211, USA.}
\thanks{corresponding author.}
\date{17 April 2013}

\begin{abstract}
We report on the generation of a Raman cascade spanning the wavelength range of
523 to 1750~nm wavelength range, in a
standard telecommunication graded-index multimode optical fiber.
Despite the highly multimode nature of the pump, the Raman peaks are
generated in specific modes of the fiber, confirming substantial beam cleanup
during the stimulated Raman scattering process.
\end{abstract}

\maketitle

Stimulated Raman scattering (SRS) is a well-known nonlinear process with numerous
applications including optical amplifiers, tunable lasers, spectroscopy, meteorology,
and optical coherence tomography.
SRS was first observed in silica glass fibers in 1972 by Stolen et al.~\cite{Stolen1972}.
They used a frequency-doubled, pulsed, Nd:YAG laser operating at the 532~nm wavelength
to pump a single-mode optical fiber and observed Stokes emission at 545~nm.
Since then, several groups have reported the observation of SRS in optical fibers using
various configurations~\cite{Cohen1978,Rosman,Grigor,Couny}.
Cohen and Lin~\cite{Cohen1978} generated 6 cascaded
Raman peaks in a silica fiber, pumped by a mode-locked, Q-switched, Nd:YAG laser operating
at 1064~nm.
Rosman~\cite{Rosman} observed 15 orders of cascaded Raman peaks by pumping a silica fiber
at 532~nm with a frequency-doubled Nd:YAG laser.
Other configurations involving unconventional fibers have been used for extreme
Raman-comb generation.
For example, Couny et al.~\cite{Couny} demonstrated the generation of a Raman comb spanning
wavelengths from 325~nm to 2300~nm in a 1m-long hydrogen-filled hollow-core photonic crystal fiber.

In this article, we report on the generation of new wavelengths, mediated by the SRS process,
in a standard graded-index multimode fiber (GIMF). The GIMF is pumped
at a wavelength of 523~nm, and a cascade of optical-frequency Raman peaks is generated on the Stokes side of
the pump. At high power levels the measured spectrum extends up to 1750~nm, which is the upper
detection limit of our optical spectrum analyzer (OSA).
The generation of such a wide wavelength range, extending from 523 to 1750~nm, using
a large-core telecommunication-grade multimode fiber distinguishes our results from those carried
out in small-core or highly customized optical fibers.

Multimode optical fibers are easy to handle and are also easy to align to external sources; however,
their large core diameter is perceived as undesirable for nonlinear applications. Despite a lower
effective nonlinearity associated with a larger core of conventional multimode fibers, the multimode
nature of these fibers can play an important role in some nonlinear applications. In particular,
the presence of multiple propagating modes with different dispersive properties results in expanded
phase-matching opportunities for the generation of four-wave mixing (FWM) signals in optical fibers~\cite{Grigor,AgrawalBook,StolenFWM,XuFWM}.
The GIMF used in our experiments has two desirable properties that make it particularly suitable for SRS
generation. First, because the effective modal area of each mode in the GIMF scales only as square root
of the core area, the effective nonlinearity of some propagating modes is comparable with conventional
single-mode fibers~\cite{MafiNL}. Second, a relatively high ${\rm GeO_2}$ content in the core of the
standard telecommunication GIMF used in our experiments results in a higher peak Raman gain coefficient
compared with silica-core fibers~\cite{Polley}.

The pump used in our experiment is a frequency-doubled, Q-switched, Nd:YLF laser operating at 523~nm wavelength,
and its 8-ns duration pulses are coupled to the input tip of the fiber using a microscope objective. The laser beam
is not diffraction-limited, and many modes in the fiber are excited simultaneously. The GIMF is a 1-km long standard,
50/125-$\mu$m, bare fiber (Corning ClearCurve OM2).
The output spectrum is measured using a CCS200 spectrometer (from Thorlabs) operating in the range of 200--1000~nm and
a MS9740A OSA (from Anritsu) covering the 600--1750~nm wavelength range.
We measured the energy of input pulses required for reaching the Raman threshold immediately after coupling
into the fiber (less then 1-meter of propagation inside the GIMF) to be 20.9~$\mu$J; we estimate the peak
power to be about 2.5~kW. The pulse energy decreased to about 0.515~$\mu$J
after 1-km of propagation, which is consistent with the expected attenuation of about 16 dB/km at the pump wavelength.
The measured threshold power for the first Raman peak ($P^{cr}_0$) is consistent with
$g_R P^{cr}_0 L_{\rm eff} \approx 16 A_{\rm eff}$~\cite{AgrawalBook} using the
effective length of $L_{\rm eff}=270$~m (considering the 16 dB/km attenuation), $g_R\approx 2.9\times 10^{-13}$~m/W (at 523~nm wavelength
and considering the ${\rm GeO_2}$ doping),
and an effective area of $A_{\rm eff}=\pi\times (62.5~{\rm \mu m})^2$ (considering the heavily multimode and overfilling nature of the pump).

\begin{figure}[tb!]
\centering
\includegraphics[width=60mm]{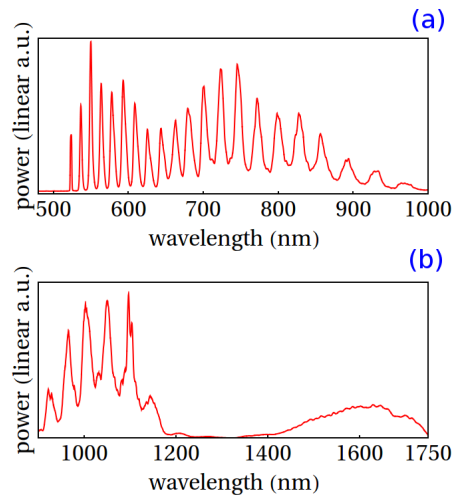}
\caption{(a) Cascaded Raman peaks measured with the spectrometer. (b)
Spectrum measured by the OSA when pump power is increased to just below the
burning threshold of fiber's tip; the spectral dip at around 1300~nm and the broad peak
beyond 1400~nm are two notable features in this infrared range.}
\label{fig:Raman_below_above_1000nm}
\end{figure}

The sequential generation of cascaded Raman peaks is initiated by the pump at 523~nm. As the pump power is increased, the first
Stokes line extracts power from the pump until it becomes strong enough to seed the generation
of next Stokes line. This process continues and more and more Raman peaks are generated with increasing pump power.
The 20 cascaded peaks shown
in Fig.~\ref{fig:Raman_below_above_1000nm}(a) extend from 523~nm to just above 1000~nm in wavelength. The estimated
input peak power of our pump pulses is 22~kW for this figure.

\begin{figure}[tb!]
\centering
\includegraphics[width=60mm]{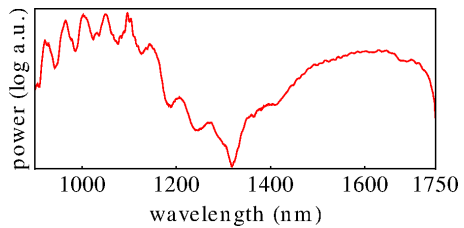}
\caption{Same as Fig.~\ref{fig:Raman_below_above_1000nm}(b) but data plotted using a logarithmic vertical scale.
The input pump power is just below the burning threshold of fiber's tip.}
\label{fig:Raman_above_900nm_log}
\end{figure}

As the pump power is further increased, even more cascaded Raman peaks appear beyond the 1000~nm wavelength range of our
spectrometer. Figure \ref{fig:Raman_below_above_1000nm}(b) shows the spectrum measured with the Anritsu OSA in the range of
900--1750~nm at the maximum pump power level (just below the burning threshold of fiber's input tip).
We stress that the two plots in Figs.~\ref{fig:Raman_below_above_1000nm}(a) and~\ref{fig:Raman_below_above_1000nm}(b)
should not be compared directly because they correspond to
different power levels and employ different vertical scales.
The spectral dip at around 1300~nm and the broad peak beyond 1400~nm are two notable features in this infrared range.
The appearance of the spectral dip centered at the 1320-nm wavelength is related to a
reduction in the SRS gain occurring near the zero-dispersion-wavelength (ZDW) of the GIMF,
where the SRS gain is suppressed due to a near-perfect phase-matching of the
FWM process~\cite{Golovchenko,Vanholsbeeck}. The dip at 1320~nm can be seen more clearly
in Fig.~\ref{fig:Raman_above_900nm_log}, where we plot the
data in Fig.~\ref{fig:Raman_below_above_1000nm}(b) on a logarithmic power scale.

In our opinion, the broad
peak centered at 1600~nm results from the onset of modulation instability in the presence of anomalous dispersion. The
resulting short pulses can experience Raman-induced spectral shifts as well as collision-based
spectral broadening, resulting in a broad supercontinuum-like feature~\cite{Chapman,Dudley}.
The generation of the longer wavelengths beyond the ZDW is a very complex phenomenon and is
heavily influenced by parametric processes. Even in the absence of perfect phase matching, FWM can seed higher-order Raman waves
that are subsequently amplified through SRS~\cite{Vanholsbeeck2,Sylvestre}.

\begin{figure}[tb!]
\centering
\includegraphics[width=60mm]{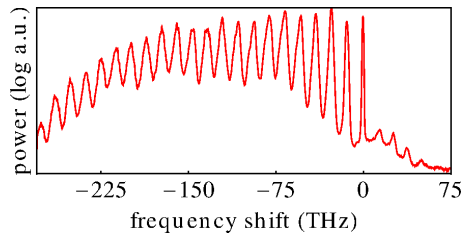}
\caption{Cascaded Raman peaks measured with the spectrometer and plotted using frequency shift
on the horizontal axis. Notice the presence of FWM peaks on the anti-Stokes side.}
\label{fig:Raman-FWM-log}
\end{figure}

The presence of efficient FWM phase-matching opportunities in the GIMF impacts considerably the generation of
the cascaded Raman peaks.
Figure \ref{fig:Raman-FWM-log} shows the spectrometer data in the frequency domain,
by plotting on the horizontal axis the frequency shift of the Raman comb relative to the pump frequency, and
power on a logarithmic scale on the vertical axis. Equal spacing of various peaks on the Stokes side is expected
for a cascaded Raman process. However, the presence of FWM frequencies on the anti-Stokes side of the pump is
the most notable feature in this figure. The phase-matched frequency counterparts of these FWM idlers on the
Stokes side can affect the location and amplitude of the cascaded Raman peaks. We observed that the strength
of the FWM signal depended on launch conditions, and FWM was absent (or highly suppressed) in some of our measurements.
{The highly multimode nature of the pump made it very difficult to find the optimum launch position for the generation
of the anti-Stokes peaks; we needed to scan the input pump beam over the fiber core to find the point at which the anti-Stokes
peaks appeared with considerable power.
}

\begin{figure}[tb!]
\centering
\includegraphics[width=60mm]{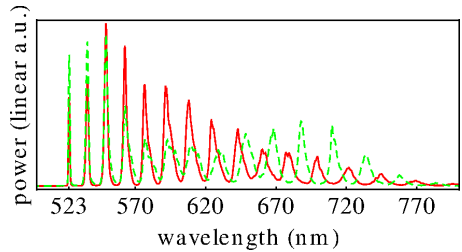}
\caption{SRS spectra measured before (red solid line) and after (dashed green line) a slight offset of the pump beam from the fiber center. Shift of Raman peaks is caused by different FWM conditions caused by excitation of different fiber modes.}
\label{fig:Raman-Shift-due-to-FWM}
\end{figure}

In order to explore the effect {of FWM on the SRS peaks}, we
{
offset the input pump beam by $\approx 15~\mu m$ from the center of the fiber core.
}
The pump now excites the GIMF modes with different power
ratios, resulting in efficient FWM in a different set of phase-matched wavelengths. The result is a
shift in the position of the spectral combs. The red and green spectra in Fig.~\ref{fig:Raman-Shift-due-to-FWM}
are measured before and after offsetting the pump laser, respectively. The shift seems to be seeded around the
third cascaded Raman peak, separated by about 50~THz from the pump frequency, which is also consistent with the
location of a FWM peak in Fig.~\ref{fig:Raman-FWM-log}.
{
We observed that the shift is reversed if the input pump beam is aligned back with the center of the fiber core.
}
Similar observations of the effect of the FWM
processes on shifting the SRS spectrum have been reported by Sharma et al.~\cite{Sharma}; they have shown that the cascaded
Raman peaks can shift depending on which modes are excited by the pump laser.

\begin{figure}[tb!]
\centering
\includegraphics[width=60mm]{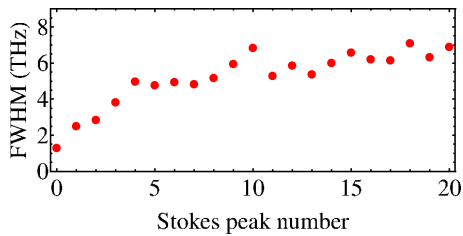}
\caption{The FWHM spectral width of the Raman peaks
from Figs.~\ref{fig:Raman_below_above_1000nm}(a) and~\ref{fig:Raman-FWM-log} are plotted
as a function of the Stokes peak number. The Stokes peak numbers 0 and 1 correspond to the pump and the first
order Raman peak, respectively.}
\label{fig:Peak-Width}
\end{figure}

In Fig.~\ref{fig:Peak-Width}, we plot the spectral width (full width at half maximum or FWHM) of the Raman peaks
from Figs.~\ref{fig:Raman_below_above_1000nm}(a) and~\ref{fig:Raman-FWM-log}
as a function of the Stokes peak number. The numbers 0 and 1 correspond to the pump and the first
order Raman peak, respectively. Ref.~\cite{Ilev} predicted that spectral width of each Raman peak
is nearly twice that of the preceding order because of the broad bandwidth of the Raman gain in fused silica.
The measured FWHM spectral widths of the pump and the first order Raman peak are 1.30~THz and 2.51~THz, respectively.
{
According to Refs.~\cite{Gersten,Ilev}, the spectral bandwidth of the Raman peak under certain conditions is nearly twice that of the pump. However, their analysis makes the undepleted-pump approximation and assumes an unchirped Gaussian profile for the pump pulse, neither of which apply to our experiment. Moreover, the noisy nature of the SRS peaks makes the analysis of Refs.~\cite{Gersten,Ilev} even less applicable to the spectrum of higher-order Raman peaks. In our case, the bandwidths of higher-order Raman peaks show an increasing trend as a function of their Stokes peak numbers.}

\begin{figure}[tb!]
\centering
\includegraphics[width=75mm]{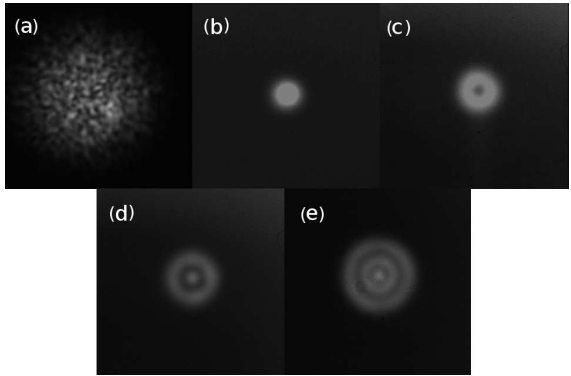}
\caption{Measured spatial profiles using a CCD-based beam profiler. Image (a) is measured with no filters. The other 4 images are
obtained using color filters centered at (b) 610~nm, (c) 700~nm, (d) 770~nm, and (e) 890~nm.}
\label{fig:modeprofiles}
\end{figure}

We also measured the transverse intensity profile of the output beam by a CCD camera using several bandpass color
filters with a FWHM spectral bandwidth of about 10~nm. The results are shown in Fig.~\ref{fig:modeprofiles}.
The profile in Fig.~\ref{fig:modeprofiles}(a) is measured without a color filter, and the interference of multiple modes can be clearly
observed as a speckle pattern. When a color filter centered at 610~nm is placed in front of the beam Fig.~\ref{fig:modeprofiles}(b), we observe a narrow
round spot that appears to correspond to the spatial profile of the fundamental ${\rm LG_{00}}$ mode of our GIMF.
We note that ${\rm LG}$ stands for Laguerre-Gaussian which are the eigenmodes of the GIMF under the weak-guidance approximation.
The ${\rm LG_{nm}}$ modes can be related to the familiar notation of the ${\rm LP_{m,n+1}}$ modes commonly used for step-index fibers.
A donut-shape spot Fig.~\ref{fig:modeprofiles}(c) is observed when a color filter centered at 700~nm is placed in front of the beam; the shape and the size of the beam
makes us believe that it corresponds to the LG$_{01}$ mode of our fiber.
We note that in practice, the two fold degeneracy of the LG$_{01}$ mode is slightly broken to orthogonal double-lobed spatial profiles
of Hermite-Gaussian modes~\cite{Kahn} due to birefringence (similar to the two polarization of the ${\rm LP_{01}}$ modes). The donut shape
in our measurement arises when both double-lobed spatial profiles are present simultaneously, primarily due to the large bandwidth (10~nm) of
the color filter used for imaging the modes resulting from the incoherent combination of the orthogonal double-lobed spatial profiles.

Spatial beam profiles corresponding to higher-order fiber modes were also seen in our experiment. As two examples, images (d) and (e)
of Fig.~\ref{fig:modeprofiles} show profiles corresponding to the ${\rm LG_{10}}$ and ${\rm LG_{20}}$ modes. They were obtained by using
optical filters with a 10-nm passband centered at 770~nm and 890~nm wavelengths, respectively. The most notable feature we want to stress
is that a GIMF can be used as a device that not only shifts the pump wavelength toward the red side through SRS but also performs the beam
cleanup owing to the fact that different-order Raman peaks generally propagate in different modes of the fiber. The main reason for the
Raman beam cleanup is that the lower order modes generally have a larger Raman gain because of their greater overlap with the higher concentration
of ${\rm GeO_2}$ near the center of the GIMF core~\cite{Polley}. A detailed analysis of SRS-induced beam cleanup in graded-index multimode optical
fibers can be found in Ref.~\cite{Terry}.
Chiang reported similar results for higher order SRS combs in a 30-m-long fiber~\cite{Chiang}. However, only ${\rm LP_{01}}$ mode (corresponding
to ${\rm LG_{00}}$ mode here) was observed for a 1-km-long fiber. In our experiments, we observed higher-order modes even for a 1-km-long fiber, and
beam cleanup was not at the same level reported in Ref.~\cite{Chiang}.

In conclusion,
we have used a standard, telecommunication-grade, graded-index multimode fiber for SRS generation by pumping it at 523~nm with 8-ns pulses.
We observe multiple cascaded Raman peaks extending up to 1300~nm. Beyond that wavelength, the nature of dispersion changes from normal
to anomalous because our fiber has its zero-dispersion wavelength near 1320~nm. At higher pump powers, in addition to the multiple cascaded
Raman peaks, we observe a single broadband spectral peak, extending from 1350 to 1750~nm. Its origin appears to be related to the
formation of solitons through modulation instability, intrapulse raman scattering, as well as collision-based spectral broadening.
Such features have been observed in the past for single-mode fibers
or few-mode step-index fibers (see, e.g., Mussot et al.\cite{Mussot}).
Our experiments show that a supercontinuum can also form in a
highly multimode telecommunication-grade, graded-index multimode fiber.
The multimode nature of the fiber can also be useful from a practical standpoint. For example, we observed that different spectral peaks have spatial
patterns that correspond to different fiber modes. This feature can be useful for beam cleanup. Future efforts will focus on extending the spectrum to the
infrared region and on stabilizing the frequency and power of individual comb lines for practical applications.

\section*{Acknowledgments}
A. Mafi acknowledges support from the UWM Research Growth Initiative grant
and the Air Force Office of Scientific Research under Grant FA9550-12-1-0329.


\end{document}